\documentclass{article}
\usepackage{setspace}
\usepackage{amsmath}
\usepackage{amssymb}
\usepackage[utf8]{inputenc}
\usepackage[margin=1.25 in]{geometry}
\usepackage{setspace}
\usepackage{tikz}
\usepackage{rotating}
\usepackage{csquotes}
\usepackage{booktabs}
\usepackage{caption}
\usepackage{subcaption}
\usepackage{graphicx}
\usepackage[multiple]{footmisc}
\usepackage[flushleft]{threeparttable}
\usepackage{hyperref}

\title{The 2020 Sturgis Motorcycle Rally and COVID-19\thanks{We would like to thank SafeGraph for providing data.}}

\author{Yong Cai\thanks{Northwestern University. Email: yongcai@u.northwestern.edu} \\ 
\and Grant Goehring\thanks{Boston University.  Email: grantg@bu.edu }}

\date{September 5, 2020}
\begin{document}
\maketitle

\begin{abstract}
 \noindent The Sturgis Motorcycle Rally that took place from August 7-16 was one of the largest public gatherings since the start of the COVID-19 outbreak. Over 460,000 visitors from across the United States travelled to Sturgis, South Dakota to attend the ten day event. Using anonymous cell phone tracking data we identify the home counties of visitors to the rally and examine the impact of the rally on the spread of COVID-19. Our baseline estimate suggests a one standard deviation increase in Sturgis attendance increased COVID-19 case growth by 1.1pp in the weeks after the rally. 
    \end{abstract}

\newpage

\begin{doublespacing}

\section{Introduction}
In response to the COVID-19 pandemic leaders have adopted a variety of policies to stop the virus from spreading. Restricting large gatherings has been one such policy recommended by the Center for Disease Control (CDC) and adopted by many state and local governments to help reduce transmission.\footnote{ CDC guidance on in-person gatherings can be found here: \url{https://www.cdc.gov/coronavirus/2019-ncov/community/large-events/considerations-for-events-gatherings.html}} The annual Sturgis Motorcycle Rally that took place in Sturgis, South Dakota from August 7-16 was one of the largest public gatherings in the United States since the start of the pandemic. Over 460,000 people from across the United States travelled to Sturgis to participate in the ten day event.\footnote{ https://dot.sd.gov/transportation/highways/traffic} This paper studies the effect of the Sturgis rally on subsequent COVID-19 case growth in the home counties of rally attendees.

\indent This study uses anonymous cell phone tracking data from Safegraph to identify individuals that were in Sturgis during the rally period. The data also provide information on the individual's home census block group allowing us to identify areas that had relatively more rally attendees. Combining this information with county-level COVID-19 case data from the \textit{New York Times} we find that counties with relatively more rally attendees have higher COVID-19 case growth rates in the weeks following the rally. 

We contribute to the rapidly growing body of work on the efficacy of social distancing and related policies. Whereas many papers have found that social distancing measures were successful in containing the spread of COVID-19 (e.g. Andersen, 2020; Courtmanche, Garuccio, Le, Pinkston and Yelowitz, 2020), studies of large-scale public gatherings have yielded seemingly contradictory results. For instance, Dave et. al (2020) study whether the Black Lives Matter protests in the aftermath of George Floyd's death on May 25, 2020 increased COVID-19 cases. By comparing cities where protests did or did not occur as well as variation in the start dates they find that protests did not increase COVID-19 case growth rate. They find evidence that a greater share of people stayed home due to the resulting curfews and general unrest, possibly offsetting the effect of the protests. A synthetic control study on President Trump's political rally in Tulsa, Oklahoma on June 20, 2020 also found no effect on COVID-19 case growth (Dave, Friedson, Matsuzawa, McNichols, Redpath, Sabia, 2020). The authors suggest that offsetting behavior, as well as smaller-than-expected crowd attendance might be important in limiting the effects of the rally.

There are several possible reasons why COVID-19 cases increased after the Sturgis rally. First, the Sturgis rally was much larger and over a longer duration than these other events. Second, while the previously cited studies identify offsetting behavior during large public gatherings we find no such attenuating behavior in counties with more Sturgis rally attendees. This suggests people were not aware of Sturgis attendees in their communities and therefore did not change their behavior in response.

\indent The rest of the paper will proceed as follows. Section 2 will provide background on the Sturgis rally and introduce the data used to identify Sturgis attendees. Section 3 will introduce our empirical strategy and present the results. Section 4 will conclude.

\section{Background and Data}
The Sturgis Motorcycle Rally is an annual motorcycle rally held in Sturgis, South Dakota. Since 2011 the South Dakota Department of Transportation (DOT) has collected data on the total number of vehicles entering Sturgis during the rally. In 2020 DOT tracked over 460,000 vehicles entering the rally compared to a ten year average of 547,882. Rally attendance outperformed many early media expectations predicting the rally to be half the normal size due to the pandemic.\footnote{See for instance \url{https://www.huffpost.com/entry/sturgis-motorcycle-rally-south-dakota-covid_n_5f2984afc5b6a34284bfed0b}} Rally attendees overshadow the native Sturgis population of 7,000, and are an important source of business for local establishments. In June, the Sturgis City Council voted 8-1 to allow the rally to proceed, and the Governor of South Dakota, Kristi Noem, enthusiastically urged that the rally take place as usual.\footnote{ On August 7, 2020 Governor Noem tweeted: ``\#Sturgis2020 kicks off today. Welcome to South Dakota! ... We've been `Back to Normal' for over 3 months, and South Dakota is in a good spot."} While motorcycle riding is a common interest among attendees, the rally in many respects functions as a large multi-day party. Concerts are held and attendees frequent local bars and drink in camp sites set up for the rally.\newline
\indent We use cell phone tracking data from SafeGraph to identify rally attendees and their home location. For a nationally representative sample of cell phones the data track the census block group containing the device as well as its ``home" census block group. SafeGraph determines the home location of the cell phone by identifying the census block group where the cell phone is most frequently located at night over a six week period.\footnote{SafeGraph has made the data available to COVID-19 researchers here: \url{https://docs.safegraph.com/docs/social-distancing-metrics}} Figure 1 shows the distribution of Sturgis exposure at the county-level. The map displays the proportion of a county's devices located in Sturgis during the rally period from August 7-16. Most of the rally attendees are from neighboring Great Plains states as well as from the Midwest and western United States. We combine the SafeGraph data with county-level COVID-19 case data compiled by the New York Times and more detailed state-level COVID-19 data from the COVID Tracking Project.\footnote{The county-level data from the New York Times is available here: \url{https://github.com/nytimes/covid-19-data}}\footnote{ The data from the COVID Tracking Project is available here: \url{https://covidtracking.com/data}}

\begin{figure}[htb!]
        \caption{Proportion of Devices at the Sturgis Rally (August 7-16, 2020)} \label{fig:timing1}
        \centering
        \includegraphics[width=\linewidth]{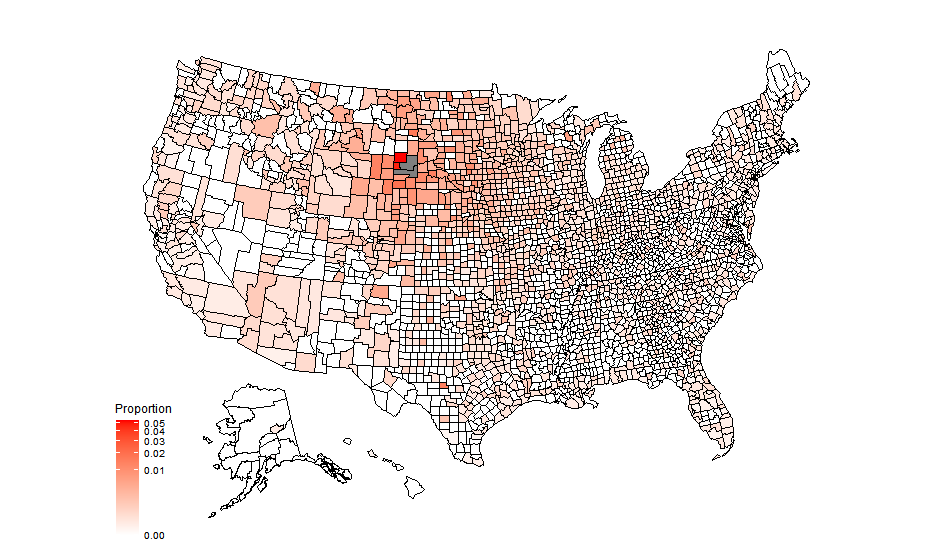} 
\end{figure}

\section{Method and Results}
We study effect of Sturgis on the spread of COVID-19, via event study and difference-in-differences.

\subsection{Event Study}

The event study is motivated by identification concerns. In particular, Sturgis exposure might be correlated with individual attitudes towards COVID-19, as well as the level of local government response (see Painter and Qiu, 2020; Allcott, Boxell, Conway, Gentzkow, Thaler and Yang, 2020; among others). This would lead to positive bias in our estimates. Conversely, individuals from counties with fewer COVID-19 cases might have subjective beliefs that underweight the probability of contracting the illness, making them more likely to attend the Rally. 

The event study takes the form:
\begin{equation*}
    Y_{c,t} = \sum_{\tau \,=\, 06/07}^{08/30} \beta_\tau \cdot {Sturgis}_{c}\cdot\mathbf{1}{\{t = \tau\}} +  X_{c,t}'\gamma + \varepsilon_{c,t}~.
\end{equation*}
where $Y_{c,t}$ is the growth rate of COVID-19, defined as the log difference in cumulative COVID-19 cases, in county $c$ at week $t$. Given the varying norms around case reporting on the weekends, we aggregate our data to the week level. The period under consideration begins on the first week of June (week ending on June 7) and ends on the last week of August (week ending on August 30). The Sturgis Rally ran from August 7-16. We take the week ending August 16 as the period of treatment and exclude it from our analysis.\footnote{The week ending August 9 is considered to be a pre-treatment period since visitors would not have returned to their respective communities at this point in time.} We drop observations from the county containing Sturgis (Meade County) as well as adjacent counties since travel patterns between these counties are likely not due to rally attendance. While we do not report the specifications, results are robust to different sample selections.\footnote{We exclude: Perkins, Butte, Meade, Ziebach, Lawrence, and Pennington Counties. Results are also qualitatively similar when all counties in South Dakota or only Meade and Pennington Counties, which contain Sturgis and Rapid City, are removed.}

${Sturgis}_{c}$ is one of two measures of county exposure to the Sturgis Rally. Our preferred measure is $SturgisProp$ -- the proportion of the county that visited Sturgis. We also consider $SturgisTopHalf$ -- an indicator for being in the top 50\% of $SturgisProp$. $X_{c,t}$ is a set of conditioning variables which includes lagged median percentage time spent at home, growth in state COVID-19 testing, as well as county and week fixed effects. 

We plot the $\beta_\tau$'s as well as their 95\% confidence intervals for the two types of exposure measure in the figure below: 
\begin{figure}[htb!]
    \centering
    \caption{Effect of Sturgis Exposure}
    \begin{subfigure}[t]{0.49\textwidth}
        \centering
        \includegraphics[width=\linewidth]{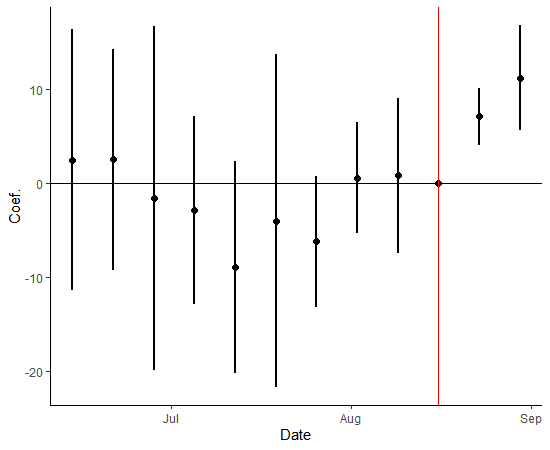} 
        \caption{$SturgisProp$: Proportion of county that visited Sturgis} \label{fig:exposure1}
    \end{subfigure}
    \hfill
    \begin{subfigure}[t]{0.49\textwidth}
        \centering
        \includegraphics[width=\linewidth]{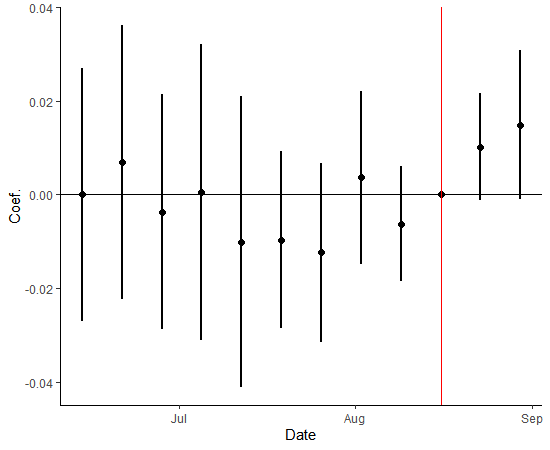} 
        \caption{$SturgisTopHalf$: Indicator for being in the top 50\% of $SturgisProp$} \label{fig:exposure2}
    \end{subfigure}
\end{figure}
Before August 17, the coefficients are generally negative and not significantly different from $0$. This suggests that there is little positive selection into Sturgis attendance. After August 17, the coefficients turn positive and significant, and appear to be increasing over time. According to the estimates, a 1pp increase in $SturgisProp$ leads to a 7pp increase in the growth rate of COVID-19 cases in the week ending August 23. This effect increases to 11pp in the week ending August 30. Put differently, a one standard deviation increase in $SturgisProp$ increases COVID-19 case growth by 1.02pp in the week immediately following the event and 1.48pp in the week after. In the same vein, a county in the top half of Sturgis exposure experiences growth in COVID-19 cases that is higher by 1.02pp relative to a county in the bottom half in the week ending August 23, rising to 1.48pp in the week ending August 30. The effect is more precisely estimated for $SturgisProp$ than $SturgisTopHalf$, possibly because the former is a more accurate measure of Sturgis exposure. 

\subsection{Difference-in-Differences}

Next, we consider the difference-in-differences (DiD):
\begin{equation*}
    Y_{c,t} = \beta \cdot {Sturgis}_{c}\cdot\mathbf{1}{\{t > 08/17\}} +  X_{c,t}'\gamma + \varepsilon_{c,t}~.
\end{equation*}
$Y_{c,t}$ and $Sturgis_{c}$ are defined as before. We consider various specifications, as explained below. Results can be found in table \ref{tab:did}. 

\begin{table}[htpb]
  \centering
\begin{threeparttable}
  \caption{Results for Difference-in-Differences}
    \begin{tabular}{lcccccc}  \toprule
          & (1)   & (2)   & (3)   & (4)   & (5)   & (6) \\
    \midrule
    Coef. & 11.014 & 0.016 & 12.546 & 14.654 & 9.591 & 2.083 \\
    S.E.  & (2.083) & (0.007) & (1.479) & (1.857) & (2.401) & (57.829) \\
    p-value & 0.000 & 0.030 & 0.000 & 0.000 & 0.000 & 0.971 \\
    \midrule
    Treatment & Prop & TopHalf & Prop & Prop & Prop & Prop \\
    County Time Trend & No    & No    & Yes   & Yes   & Yes   & Yes \\
    Exclude College & No    & No    & No    & Yes   & No    & No \\
    Balanced Panel & No    & No    & No    & No    & Yes   & No \\
    \midrule
    $N$ & 36,749    & 36,749    & 36,749    & 16,944    & 35,760   & 36,749 \\
    $R^2$ & 0.259 & 0.259 & 0.396 & 0.338 & 0.394 & 0.870 \\
    \midrule
    \end{tabular}%
    \begin{tablenotes}
        \item \leavevmode \kern-\scriptspace \kern-\labelsep 
        Notes: Standard errors clustered by state. The outcome variable is COVID-19 case growth for (1) - (5) and median percentage time spent at home for (6). Prop is the proportion of the county that visited Sturgis. TopHalf is a dummy for top 50\% of Prop.
    \end{tablenotes}
    \label{tab:did}%
\end{threeparttable}%
\end{table}

In specifications (1) and (2), $X_{c,t}$ includes lagged median percentage time spent at home, growth in state COVID-19 testing, as well as county and week fixed effects. These specifications are therefore comparable with the event studies above. Column 1 of table \ref{tab:did} shows that a 1pp increase in Sturgis attendance increased growth in COVID-19 cases by 11pp on average in the weeks following the Rally. Equivalently, a one standard deviation increase in Sturgis attendance increased growth in COVID-19 cases by 1.1pp on average. Column 2 shows that a county in the top 50\% of Sturgis attendance has 1.6pp higher growth in COVID-19 cases relative to those outside. 

In specification (3), we repeat specification (1) including linear county level time trends. Our results remain broadly unchanged allowing us to rule out confounders that are trending linearly in the counties. 

Given concerns about the effect of college re-opening, specification (4) repeats specification (3), but drops 1,679 counties with at least 1 college. Estimates based on the remaining 1,524 counties are generally consistent with previous specifications finding significant positive effects of Sturgis exposure.\footnote{College location data can be found here: \url{https://hifld-geoplatform.opendata.arcgis.com/datasets/colleges-and-universities}}

About 7\% of counties are not present in the week of Jun 7, 2020. To explore selection into data reporting we repeat specification (3) with a balanced panel. This yields specification (5), which leads to similar results as before. 

Finally, the last specification explores whether individuals observe their neighbors attending Sturgis, and change their behavior in response. Specification (6) tests this hypothesis by using median percentage time spent at home as the outcome variable. Clearly, individuals are not adjusting their behavior. This could account for the relatively larger effect of the Sturgis Rally relative to other public events (e.g. in Dave, D. M., Friedson, A. I., Matsuzawa et al. 2020). 

\section{Conclusion}

We examine the effect of the Sturgis Rally on COVID-19 case growth in the United States. We find counties with relatively more rally attendees experienced higher COVID-19 case growth in subsequent weeks. Other studies that have found large public gatherings do not affect case growth point to offsetting behavior that reduced possible COVID-19 transmission following the event. We find that stay-at-home behavior did not change in response to Sturgis exposure suggesting individuals were likely unaware of Sturgis attendees in their communities and therefore did not take take precautions to reduce their risk of exposure following the rally.   

\vspace{6mm}

\end{doublespacing}

\section*{References}

\noindent Allcott, H., Boxell, L., Conway, J., Gentzkow, M., Thaler, M., and Yang, D. (2020). Polarization and Public Health: Partisan Differences in Social Distancing during the Coronavirus Pandemic (No. w26946). National Bureau of Economic Research. \newline

\noindent Andersen, M. (2020). Early evidence on social distancing in response to COVID-19 in the United States. Available at SSRN 3569368. \newline

\noindent Courtemanche, C., Garuccio, J., Le, A., Pinkston, J., and Yelowitz, A. (2020). Strong Social Distancing Measures In The United States Reduced The COVID-19 Growth Rate: Study evaluates the impact of social distancing measures on the growth rate of confirmed COVID-19 cases across the United States. Health Affairs, 10-1377. \newline

\noindent Dave, D. M., Friedson, A. I., Matsuzawa, K., McNichols, D., Redpath, C., and Sabia, J. J. (2020). Did President Trump’s Tulsa Rally Reignite COVID-19? Indoor Events and Offsetting Community Effects (No. w27522). National Bureau of Economic Research. \newline

\noindent Dave, D. M., Friedson, A. I., Matsuzawa, K., Sabia, J. J., and Safford, S. (2020). Black Lives Matter protests, social distancing, and COVID-19 (No. w27408). National Bureau of Economic Research. \newline

\noindent Painter, M., and Qiu, T. (2020). Political Beliefs Affect Compliance with COVID-19 Social Distancing Orders. Working Paper

\end{document}